# PDA in Action: Ten Principles for High-Quality Multi-Site Clinical Evidence Generation

Authors:
Yong Chen*, PhD[1,2,3,4,5,6]
Jiayi Tong*, PhD[1,2,7]
Yiwen Lu, BS[3,4]
Rui Duan, PhD[8]
Chongliang Luo, PhD[9]
Marc A. Suchard, PhD [10,11,12]
Patrick B. Ryan, PhD[13]
Andrew E. Williams, PhD[14]
John H. Holmes, PhD[2]
Jason H. Moore, PhD[15]
Hua Xu, PhD[16]
Yun Lu, PhD[17]
Raymond J. Carroll, PhD[18]
Scott L. Zeger, PhD [7]
George Hripcsak, PhD[19]
Martijn J. Schuemie, PhD[13]

Affiliation of the authors:
[1]The Center for Health AI and Synthesis of Evidence (CHASE), Perelman School of Medicine, The University of Pennsylvania, Philadelphia, PA, USA
[2]Department of Biostatistics, Epidemiology, and Informatics, Perelman School of Medicine, The University of Pennsylvania, Philadelphia, PA, USA
[3]The Graduate Group in Applied Mathematics and Computational Science, School of Arts and Sciences, University of Pennsylvania, Philadelphia, PA, USA
[4]Penn Institute for Biomedical Informatics (IBI), Philadelphia, PA, USA
[5]Leonard Davis Institute of Health Economics, Philadelphia, PA, USA
[6]Penn Medicine Center for Evidence-based Practice (CEP), Philadelphia, PA, USA

[7]Department of Biostatistics, Johns Hopkins Bloomberg School of Public Health, Baltimore, MD, USA
[8]Department of Biostatistics, School of Public Health, Harvard University, Boston, MA, USA
[9]Division of Public Health Sciences, Department of Surgery, Washington University in St. Louis, St. Louis, MO, USA
[10]Department of Veterans Affairs Informatics and Computing Infrastructure, Tennessee Valley Healthcare System VA, Nashville, TN, USA
[11]Department of Internal Medicine, University of Utah School of Medicine, Salt Lake City, UT, USA
[12]Department of Biostatistics, University of California, Los Angeles, CA, USA
[13]Epidemiology, Janssen Research & Development, Titusville, NJ, USA
[14]Tufts University School of Medicine, Boston, MA, USA
[15]Department of Computational Biomedicine, Cedars-Sinai Medical Center, Los Angeles, CA, USA
[16]Department of Biomedical Informatics and Data Science, Yale University, New Haven, CT, USA
[17]Center for Biologics Evaluation and Research, Food and Drug Administration, Silver Spring, MD, USA
[18]Department of Statistics, Texas A&M University, College Station, TX, USA
[19]Department of Biomedical Informatics, Columbia University, New York City, NY, USA

*Corresponding authors:

Yong Chen, PhD, ychen123@upenn.edu
Address: Blockley Hall 602, 423 Guardian Drive Philadelphia, PA 19104
Office: 215-746-8155

Jiayi Tong, PhD, jtong@jhu.edu
Address: 615 N Wolfe St, Baltimore, MD 21205

**ABSTRACT**

**Background:** Distributed Research Networks (DRNs) offer significant opportunities for collaborative multi-site research and have significantly advanced healthcare research based on clinical observational data. However, generating high-quality real-world evidence using fit-for-use data from multi-site studies faces important challenges, including biases associated with various types of heterogeneity within and across sites and data sharing difficulties. Over the last ten years, Privacy-Preserving Distributed Algorithms (PDA) have been developed and utilized in numerous national and international real-world studies spanning diverse domains, from comparative effectiveness research, target trial emulation, to healthcare delivery, policy evaluation, and system performance assessment. Despite these advances, there remains a lack of comprehensive and clear guiding principles for generating high-quality real-world evidence through collaborative studies leveraging the methods under PDA.

**Objective:** The paper aims to establish ten principles of best practice for conducting high-quality multi-site studies using PDA. These principles cover all phases of research, including study preparation, protocol development, analysis, and final reporting.

**Discussion:** The ten principles for conducting a PDA study outline a principled, efficient, and transparent framework for employing distributed learning algorithms within DRNs to generate reliable and reproducible real-world evidence.

**INTRODUCTION**

Over the past decades, there has been a substantial increase in the number of distributed research networks (DRNs) across healthcare systems, including the Observational Health Data Sciences and Informatics (OHDSI) [1], the National Patient-Centered Clinical Research Network (PCORnet) [2], the NIH-funded Health Care Systems Research Collaboratory [3,4], the Biologics Effectiveness and Safety (BEST) Initiative [5], and the Sentinel Initiative of the Food and Drug Administration (FDA) [6,7].These research networks have greatly expanded the prospects for multi-site research and public health surveillance activities across multiple data partners. In addition, the COVID-19 pandemic has further facilitated the development of various national and international multi-institutional research consortia, such as the 4CE consortium [8], the RECOVER initiative [9], N3C [10], among others. However, leveraging multi-site clinical observational data, such as electronic health records (EHRs), administrative claims, and disease registries for clinical evidence generation presents significant challenges [11]. Key barriers include restrictions on patient-level data sharing, the need for potentially iterative and resource-intensive communications across sites within the networks, the presence of various types of heterogeneity, and the reliance on labor-intensive human-in-the-loop processes. Ensuring analytic transparency and reproducibility further compounds these challenges, making multi-site research both methodologically intricate and operationally demanding.

In response to these challenges, substantial efforts have been devoted to developing federated learning algorithms for healthcare. Early examples include GLORE (Grid Binary Logistic Regression) [12], a distributed algorithm for conducting logistic regression, and WebDISCO (a web service for distributed Cox model learning), which enables fitting the Cox proportional hazards models across sites [13]. In particular, PDA (Privacy-Preserving Distributed Algorithms) [14,15] have been developed to further facilitate the use of clinical observational data from DRNs. PDA include a broad spectrum of distributed and federated learning algorithms specifically designed to address the methodological and operational complexities of multi-site studies and large-scale public health surveillance activities involving multiple data partners. Throughout this paper, we use the terms "*distributed*" and "*federated*" interchangeably to refer to settings in which patient-level data remain stored locally and are not shared across sites. The " PDA framework" refer collectively to all algorithms developed under the PDA paradigm, including those currently implemented in the PDA package [16] and future ones.

The algorithms within the PDA framework are specifically designed to protect *patients' confidentiality* by only requiring the sharing of summary statistics, not individual patient-level data, and are implemented under data usage agreements and institutional review board (IRB) protocols. These algorithms aim to achieve *statistical accuracy* equivalent to that of the ideal pooled analyses, which is considered as the gold standard when direct pooling of patient-level data is feasible [12,17]. Furthermore, the PDA framework supports a broad spectrum of analytical tasks, including comparative effectiveness research, causal inference, target trial emulation, health disparities assessment, and health policy evaluation. From a modeling standpoint, it accommodates diverse data structures and outcome types, such as binary, count, continuous, and time-to-event outcomes, along with advanced extensions such as competing risks, multivariate associations, and high-dimensional feature modeling.

What distinguishes PDA from previous efforts is its emphasis on *streamlining communication rounds*, that is the exchange of summary statistics across participating data partners. In multi-site studies operating using decentralized data networks, the most time-intensive steps involve repeated communication rounds that require coordination, synchronization, and secure data transfer across institutions. By substantially reducing the number of these exchanges, especially in modeling fitting, PDA improves the computational efficiency, lowers latency, and reduces the risk of miscommunication during summary statistics transfer. This design makes distributed algorithms more practical, scalable, and efficient for real-world healthcare and clinical research applications. Importantly, while emphasizing the minimization of communication rounds for sharing summary statistics, PDA ensures that data-sharing constraints do not compromise the accuracy and validity of the estimates.

Beyond communication efficiency, the PDA framework is specifically designed to be robust to, or explicitly account for, the potential existence of heterogeneity in treatment effects or other parameters of scientific interest. Ignoring heterogeneity within and across sites often leads to biased estimates and misleading quantification of evidence. Common types of heterogeneity encountered in DRNs include differences in data provenance, distributions, modalities, quality, and design. The current PDA framework features a comprehensive set of algorithms tailored to address or accommodate these variations.

Each algorithm within the PDA framework has been evaluated in a diverse set of studies with clinical observational data and has been widely adopted across diverse clinical fields including opioid use disorder (OUD) [18], dementia and other aging-related conditions [19], myocardial infarction [20], pediatric conditions [21], fetal loss [17,22], COVID-19 [23–25], long COVID [21], pediatric Crohn's disease (PCD) [26], health disparities and fairness [27], and health policy [28]. To date, the PDA R package has been downloaded more than 13,300 times since its release in 2020. The framework has been utilized in more than 25 national and five international studies, has contributed to more than 35 methodological papers, and is currently being applied in eight ongoing studies. Furthermore, the PDA framework has been adopted by institutions and organizations worldwide, including the International Agency for Research on Cancer (IARC) in France, Peking University Health Science Center, Ajou University Graduate School of Medicine, The Information System for Research in Primary Care (SIDIAP), Erasmus University Medical Center, and Hospital del Mar Research Institute.

Although various distributed learning algorithms have been developed and implemented under PDA, there are no comprehensive or clear guiding principles for conducting collaborative studies using these algorithms. Herein, we define a multi-site study employing any PDA methods as a *"PDA study"*. This paper outlines ten foundational principles that guide the design, execution, and reporting of PDA studies, spanning all stages from study preparation to dissemination of results. We also provide an overview of a typical PDA study to illustrate these principles in practice. Looking forward, PDA studies have the potential to expand the scope of clinical evidence generation to additional medical domains and downstream tasks, ultimately strengthening the foundation for trustworthy and generalizable real-world evidence.

## MATERIALS AND METHODS

### Ten Guiding principles

The PDA framework is strategically designed to ensure the scalability, applicability, and transparency of distributed learning algorithms for analyzing large-scale data from DRNs in the context of generating clinical evidence. The implementation of a PDA study adheres to ten foundational principles that guide investigators through essential steps and key considerations in multi-site studies — from the study preparation and protocol development to analysis and reporting. The framework provides a flexible and comprehensive set of distributed learning algorithms, each tailored to meet the practical challenges of data integration, such as reducing communication overheads, protecting patient confidentiality, and understanding and accounting for between-site heterogeneity. Users can select the algorithms that best align with their specific statistical objectives and underlying assumptions of their PDA study. In the following **Table 1**, we present the ten guiding principles for conducting a PDA study, covering all phases of research, including study preparation, protocol development, analysis, to final reporting.

| | |
|---|---|
| 1. | **Phase I: Collaboratively develop a study protocol**<br>During the protocol development stage, a PDA research protocol should include a data analysis plan with clear definitions of both the aggregation unit and distribution unit.<br>**Aim:** To determine the proper configuration of the study protocol for a PDA algorithm. |
| 2. | **Phase I: Verify fit-for-use clinical observational data at each unit prior to evidence generation.**<br>Before executing the analytical methods, PDA will conduct cohort assessment at multiple levels, such as cohort definition level and aggregation unit level, to identify potential misclassification errors in the phenotype.<br>**Aim:** To enhance the reliability of the evidence; improves the generalizability; facilitates sensitivity analysis |
| 3. | **Phase II: Protect the privacy of stakeholders including patients and data partners.**<br>Patient-level data remains confidential, and the identities of participating sites are anonymized to specific groups when necessary.<br>**Aim:** To ensure data privacy is protected at the level required across the DRN. |
| 4. | **Phase II: Ensure generation of valid evidence by leveraging diverse forms of heterogeneity.**<br>PDA is aware of the potential existence of heterogeneity from different aspects and provides strategies to account for or leverage diverse forms of heterogeneity to ensure the generation of reliable evidence.<br>**Aim:** To understand and make use of data heterogeneity. |
| 5. | **Phase II: Use reliable methods and ensure statistical accuracy, mitigating errors arising from the distributed nature of the algorithm.**<br>PDA study employs theoretically and/or empirically validated methods to guarantee statistical accuracy.<br>**Aim:** To minimize accuracy loss due to data sharing constraints. |
| 6. | **Phase II: Support scalable collaborative learning**<br>The PDA framework is designed for scalability, accommodating analyses that range from a few units to thousands, and capable of handling unit sizes as small as one or as large as millions.<br>**Aim:** To ensure the scalability of the algorithms and minimizes the interactive burden of participating sites. |
| 7. | **Phase II: Allow flexible downstream tasks, including characterization, prediction, comparative studies**<br>PDA will also offer a variety of strategic frameworks that combine model fitting with downstream analyses.<br>**Aim:** To seamlessly integrate with existing research workflows or provides a comprehensive end-to-end solution, covering all stages from data quality assessment to analysis and reporting. |
| 8. | **Phase II: Facilitate adoption of ready-to-implement functionality across institutions**<br>PDA will effectively implement distributed learning algorithms in practice by leveraging PDA-OTA websites to enrich user experience, manage collaborative studies, facilitate smooth and time-efficient communication, and enable the transfer of minimal resource requirements.<br>**Aim:** To facilitate the readiness of utilization. |
| 9. | **Phase III: Adhere to transparency to facilitate advancements in open science**<br>The analysis code, R package, and study protocol are made accessible to data partners, collaborators, and scientific community for review and evaluation.<br>**Aim:** To enhance transparency and reproducibility. |
| 10. | **Phase III: Promote collaborative interpretation and knowledge translation**<br>Once clinical evidence has been generated, the PDA promotes to stimulate discussions both within the DRN and beyond.<br>**Aim:** To enhance translation of research findings. |

*Table 1: Ten Guiding principles of implementing a PDA study. **Phase I: Preparation Phase; Phase II: Analysis Phase; Phase III: Reporting Phase***

**Overview of a principled PDA study**

**Figure 1** illustrates the overall workflow of a typical PDA study, which is organized into three phases: Preparation, Analysis, and Reporting.

In the Preparation phase, the research protocol is collaboratively defined by the participating sites, including specification the research question and identification of the aggregation and distribution units (**Principle 1**). To ensure the data are fit-for-use, potential risks of bias are assessed through quality assessment of cohort, such as cohort diagnostics [29] (**Principle 2**).

During the Analysis phase, patient-level data remain securely stored at each site to protect patient confidentiality (**Principle 3**). The selected PDA algorithm leverages heterogeneity across sites (**Principle 4**) to strengthen the robustness of evidence, meanwhile, maintaining accuracy through systematic and reproducible statistical learning methods (**Principle 5**). Sites collaborate at scale by executing analyses locally and sharing only summary statistics (**Principle 6**). The PDA framework further supports flexible downstream analytic tasks (**Principle 7**) and support implementation readiness for real-world applications (**Principle 8**).

Finally, in the Reporting phase, PDA promotes open science is promoted by publicly disseminating pre-specified protocols, analysis code, and R packages through publicly accessible repositories such as GitHub and CRAN (**Principle 9**). This approach ensures transparency and encourages widespread collaboration to critically interpret findings and assess the implications within the research community (**Principle 10**).

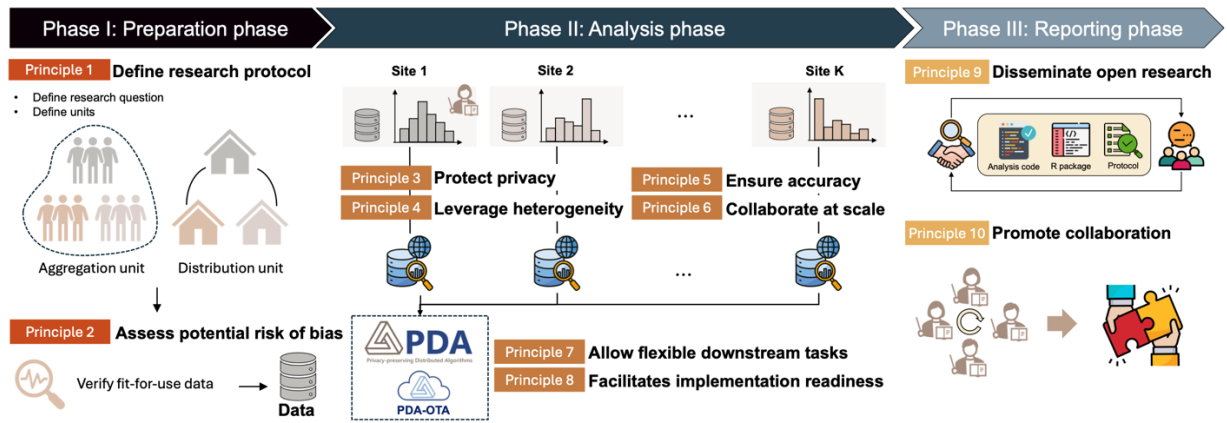

*Figure 1*: Overview of a typical PDA study and its ten guiding principles. The PDA workflow consists of three sequential phases: (I) Preparation, which includes defining the research protocol and assessing data quality and potential biases (Principles 1–2); (II) Analysis, where participating sites retain patient-level data locally and collaborate through secure exchange of summary statistics using distributed learning algorithms that leverage heterogeneity and ensure accuracy (Principles 3–8); and (III) Reporting, which emphasizes transparency, open research dissemination, and sustained collaboration (Principles 9–10).

**Preparation Phase: Study protocol design and cohort diagnostics (Principles 1 and 2)**

Before implementing statistical model fitting within the PDA framework, the protocol development stage is a crucial step in multi-site collaborative studies. During this stage, investigators develop the study analysis plan and protocol by specifying the study design, study cohort with inclusion/exclusion criteria, and selecting appropriate distributed learning algorithms. Equally important is the explicit definition of the aggregation and distribution units. The *Aggregation Unit* represents to the unit of study interest in the analytical models, whose index is included in the model to distinguish the subjects across different units. This can be an individual hospital, clinical site, or a collective network, each distinctly marked by a site/study identification number for analytical inclusion. The ***Distribution Unit*** defines the operational unit of communication, including the data partners who exchange the summary statistics for collaborative modeling tasks. The distribution unit may coincide with, or differ from, the aggregation unit depending on study configuration. The PDA framework provides flexible solutions for multi-site data integration and analysis, accommodating various configurations of aggregation and distribution units and ensuring operability and interpretability of the results.

Following the specifications of a PDA study, Principle 2 emphasizes the importance of a thorough assessment of study cohorts. This step involves confirming their suitability and ensuring that clinical observational data at each unit is fit-for-use before initiating data analysis, which is performed locally. Specifically, prior to deploying the analytical methods, the PDA framework undertakes detailed evaluation such as the cohort diagnostics [29] across multiple dimensions, such as the cohort definition level and aggregation unit level, to pinpoint and address any potential misclassification errors within the phenotype data. These diagnostics yield critical insights into the availability of cohorts across participating databases, outcome incidence rates, distributions of participant characteristics, and overlaps among cohorts, among other key quality metrics.

**Analysis Phase: Protect patient confidentiality, unfold heterogeneity, and ensure reliability (Principles 3, 4, 5)**

Within a PDA study, the framework prioritizes protection of patient confidentiality throughout the data analysis process. The PDA framework safeguards two aspects of privacy: hospital-level privacy and patient-level data privacy (Principle 3). To project the identities of hospitals from disclosure, PDA meticulously safeguards the identities of our data partners, particularly in projects where confidentiality is required. For example, in a hospital profiling project aimed at evaluating and ranking hospital performance, where institutions may be hesitant to publicize their rankings, access to sensitive information is restricted to the coordinating center. Individual sites can view only their own rankings and are not provided with the identities or performance results of other participating institutions. Regarding the protection of patient-level data, depending on the protection level required at each DRN, PDA can integrate various protection techniques such as homomorphic encryption [30–32] and differential privacy [33,34] to ensure security of the summary statistics of individual data transferred across sites, as dictated by the protocol or required by our data partners' practical needs. The summary statistics will only be used for agreed-upon purposes based on the pre-specified study protocol.

Within the scope of DRNs, the PDA framework aims to better understand, systematically account for, and leverage various types of between-site heterogeneity (Principle 4), including, but not limited to, distribution shift, variations in coding practices, intrinsic population differences, modality disparities, data quality variations, sampling mechanisms, and temporal factors as shown in **Figure 2**. The heterogeneity is typically assumed at the aggregation unit level. The goal is to produce reliable evidence that accurately represents the intricate nature of real-world healthcare scenarios, thus ensuring that the research findings are not only statistically robust but also widely applicable to a range of clinical settings.

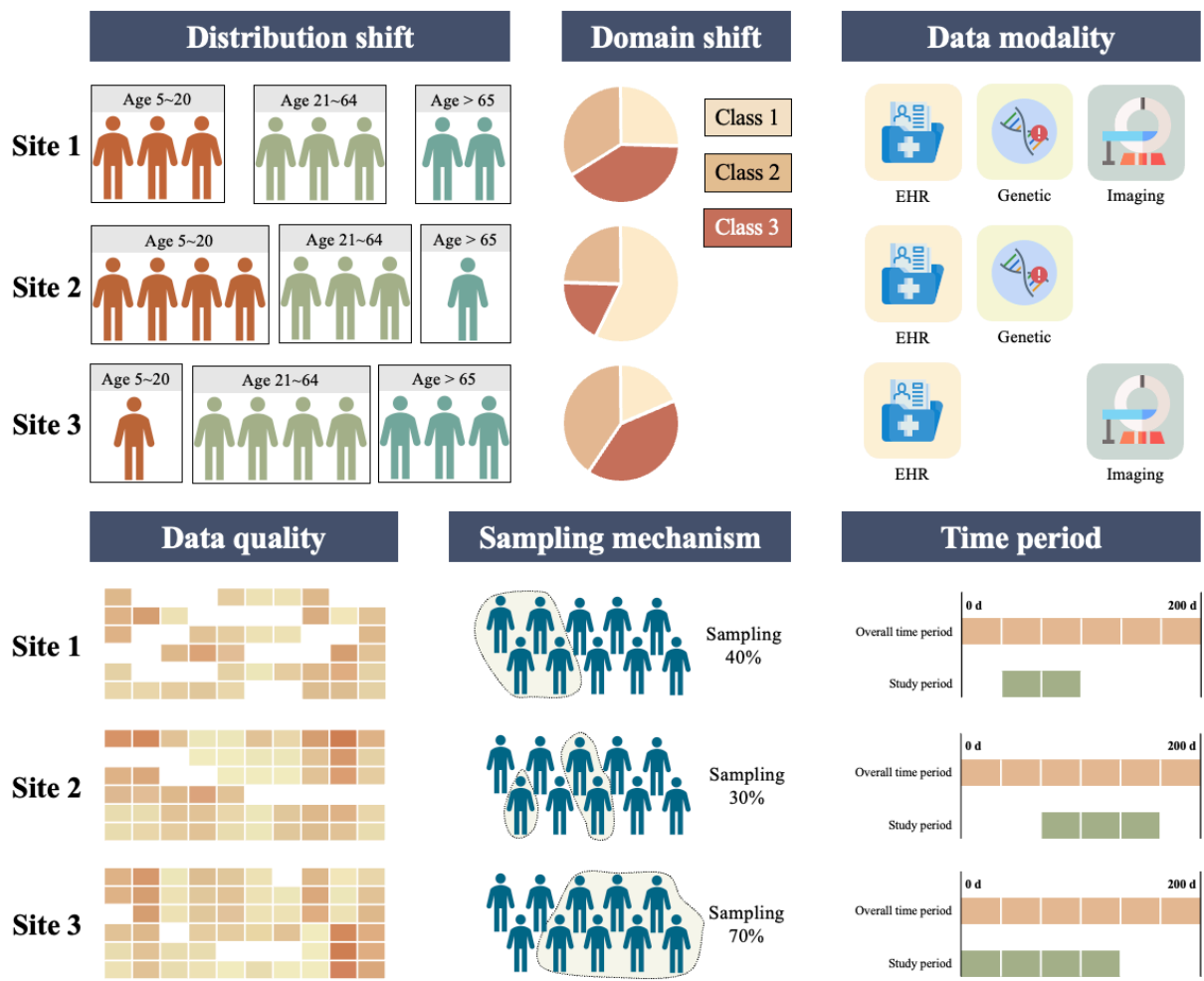

*Figure 2*: Examples of between-site heterogeneity. Between-site heterogeneity may arise from differences in population distributions, outcome domains, data modalities, data quality, sampling mechanisms, and study time periods across participating sites. Accounting for these variations is essential for generating robust and generalizable real-world evidence within the PDA framework.

Importantly, a PDA study prioritizes the use of distributed learning methods that are both theoretically and empirically validated (Principle 5). In other words, PDA emphasizes providing results that are comparable to those obtained from pooled data analyses in scenarios where data sharing is not restricted. The methodologies within the PDA framework ensure statistical precision through comprehensive empirical evaluations, benchmarked against the gold standard approach that assumes the pooling of patient-level data. These evaluations demonstrate that PDA methods can achieve nearly equivalent accuracy, with minimal loss even under strict data-sharing constraints, highlighting the robustness and effectiveness of PDA methods.

**Analysis Phase: Scalable and efficient communication rounds & Implementation readiness with workflow management (Principles 6 and 8)**

Within a PDA study, only summary-level statistics are required to be shared through an efficient communication procedure, to ensure the achievement of statistical accuracy and optimize implementation costs. The summary statistics are usually required at the aggregation unit level. Additionally, the PDA framework is designed for scalability and the minimization of iterative burden on participating sites. It is capable of handling analyses that range from just a few units to thousands and accommodating unit sizes from a single patient to one hundred million. This flexibility ensures robust performance across a diverse array of study configurations. The frequency of communication is tailored to the specific demands of each task, depending on the selected algorithm and its unique requirements. Within the PDA framework, a number of algorithms are purposefully "*light-touch*," typically completing in two communication rounds: an initialization round led by a designated lead site to coordinate the distributed analysis, followed by a synthesis round in which sites enables the synthesis of results with each site serving as the lead. Such design improves numerical stability and yields more robust estimates. When cross-site synthesis of initial values is beneficial, a brief aggregation round before the distributed analysis can further stabilize initialization [20]. Even more efficient are "*one-shot*" methods, where only a single communication round is needed, such as DLMM [23], COLA-GLM [35], and COLA-GLMM [36]. By collapsing human-in-the-loop steps and sharply reducing network traffic, these approaches lower latency and cost, preserve privacy, and enable scalable, routine analyses across large, heterogeneous clinical networks.

A key feature of PDA is that it facilitates the implementation readiness and workflow management (Principle 8). We have developed and utilized the PDA-OTA (Privacy-preserving Distributed Algorithms Over the Air) web-based interface platform [37]. This platform serves as the operational backbone, facilitating smooth and close interactions among collaborators. As illustrated in **Figure 3,** the PDA-OTA interface displays the study's operational dashboard for a project involving eight participating sites in Round 1. The platform allows sites to upload and track summary statistics, monitor participation status, and manage data submission progress in real time. By centralizing communication and data transfer, the PDA-OTA website significantly streamlines the deployment of distributed learning algorithms, ensuring efficient and effective application in diverse research contexts.

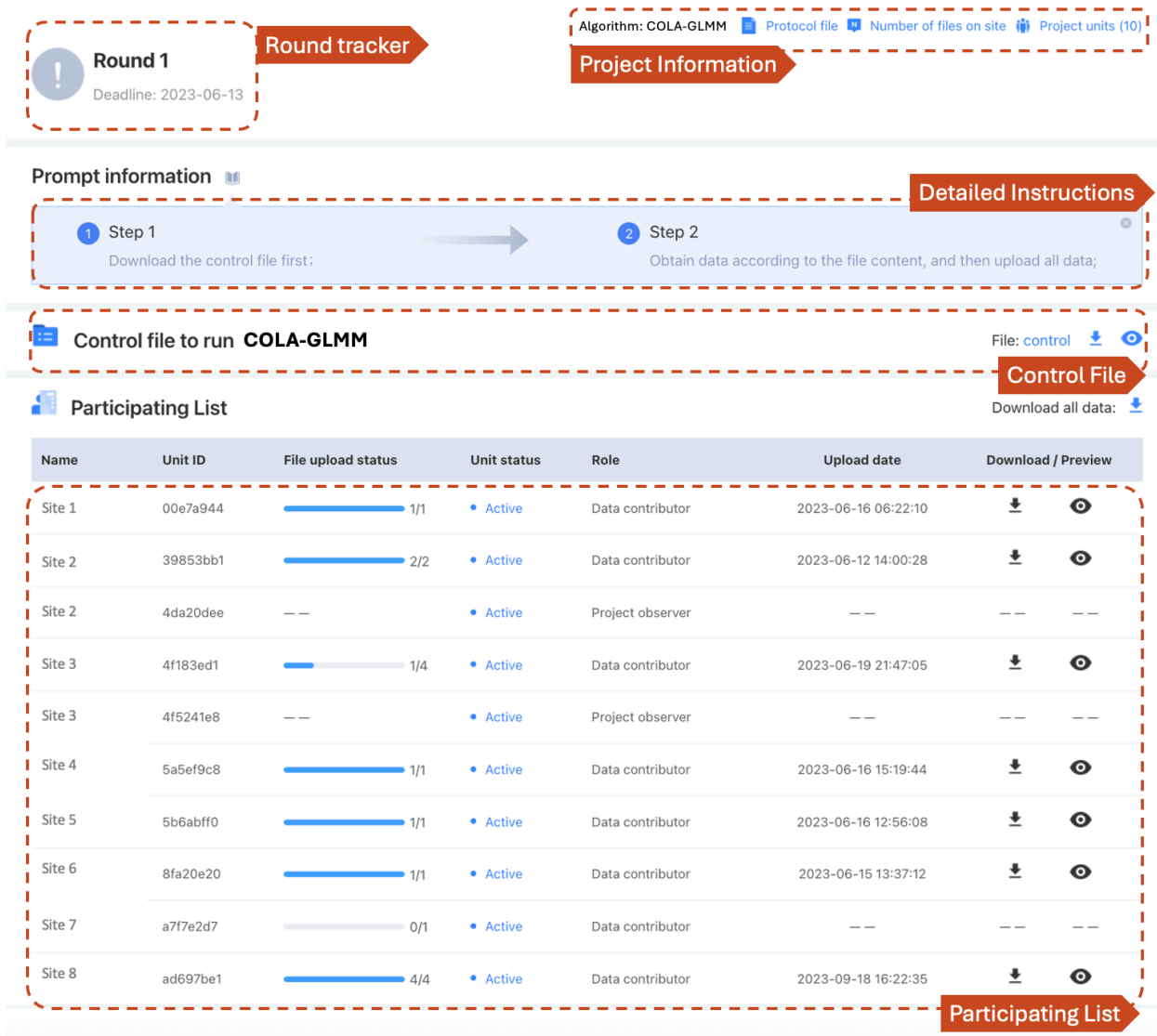

*Figure 3*: Screenshot of PDA-OTA website: an example of transferring summary statistics and monitoring project status in the context of conducting a PDA study

**Analysis Phase: Enable capacity of conducting downstream tasks (Principles 7)**

Beyond offering distributed learning analytical models and synthesizing information from multiple databases within DRNs, PDA offers users flexible downstream capabilities to extend analyses into diverse and high-impact areas of healthcare research. All downstream analyses are implemented in accordance with the guidelines outlined within the study protocol, ensuring methodological consistency and transparency. In other words,

PDA enables adaptive deployment and flexible execution of distributed learning algorithms, tailored to the specific objectives and requirements of each project.

For example, in the context of hospital profiling, which evaluates and compares healthcare delivery across hospitals, PDA supports a "Distributed Hospital Comparer" framework. This framework consists of two primary modules: a distributed learning module for fitting Generalized Linear Mixed Effects Model (GLMM), which is certified model for hospital profiling by National Quality Forum [28], and a counterfactual modeling module designed to address the patient-mix variation among different hospitals. Another example under PDA equipped with downstream analysis module is dGEM-disparity [27], a decentralized GLMM designed to quantify health disparities attributable to site-of-care differences.

In addition, PDA supports causal inference and target trial emulation by replicating clinical trial analyses using observational data, providing a scalable and effective alternative to traditional randomized controlled trials. Together, these applications demonstate PDA's capacity to deliver rigorous, policy-relevant evidence that informs healthcare quality assessment, guides interventions, and advances equity-focused decision-making.

**Reporting Phase: Transparency (Principle 9)**

The PDA framework requires the adherence to transparency principles to facilitate replication, evaluation, and community evaluation. The algorithms developed under the PDA framework are available as open-source resources through an R package and the GitHub repository [14,16]. Users are encouraged to make their analysis code and study protocols accessible to data partners, collaborators, and the broader scientific community to enable thorough review and precise reproducibility. In addition, summary statistics and analysis results are required to be made transparent among all participating units within a PDA study, reinforcing openness and scientific integrity throughout the research process.

**Reporting Phase: Promotion of a collaborative community on clinical implication and results dissemination (Principle 10)**

PDA facilitates the dissemination of analysis results, promoting a collaborative community focused on exploring clinical implications and broadening the reach of research findings. The PDA-OTA platform integrates a structured dissemination pipeline to support transparent result sharing and collaborative interpretation across distributed research teams. Within the platform [37], each project instance includes a dedicated "Final Result" module, which serves as a centralized endpoint for posting and synchronizing analysis outputs. The coordinating center or lead site can upload the final results in multiple data-interchange formats (e.g., .json, .pdf, .png, as shown in **Figure 4**), enabling both machine-readable integration and human-readable review. Once uploaded, results are automatically propagated to all participating sites, ensuring version consistency, traceability, and open access within the project network.

From a workflow perspective, the project lead initiates the publication process by compiling the aggregated results into a draft manuscript or technical report. This document undergoes collaborative review and sign-off from all participating sites. The framework also highly encourages the inclusion domain experts to contribute clinical interpretations and insights during post-analysis review. This integrated dissemination architecture not only supports transparency and reproducibility but also accelerates the translation of distributed analytical outputs into actionable clinical and policy knowledge.

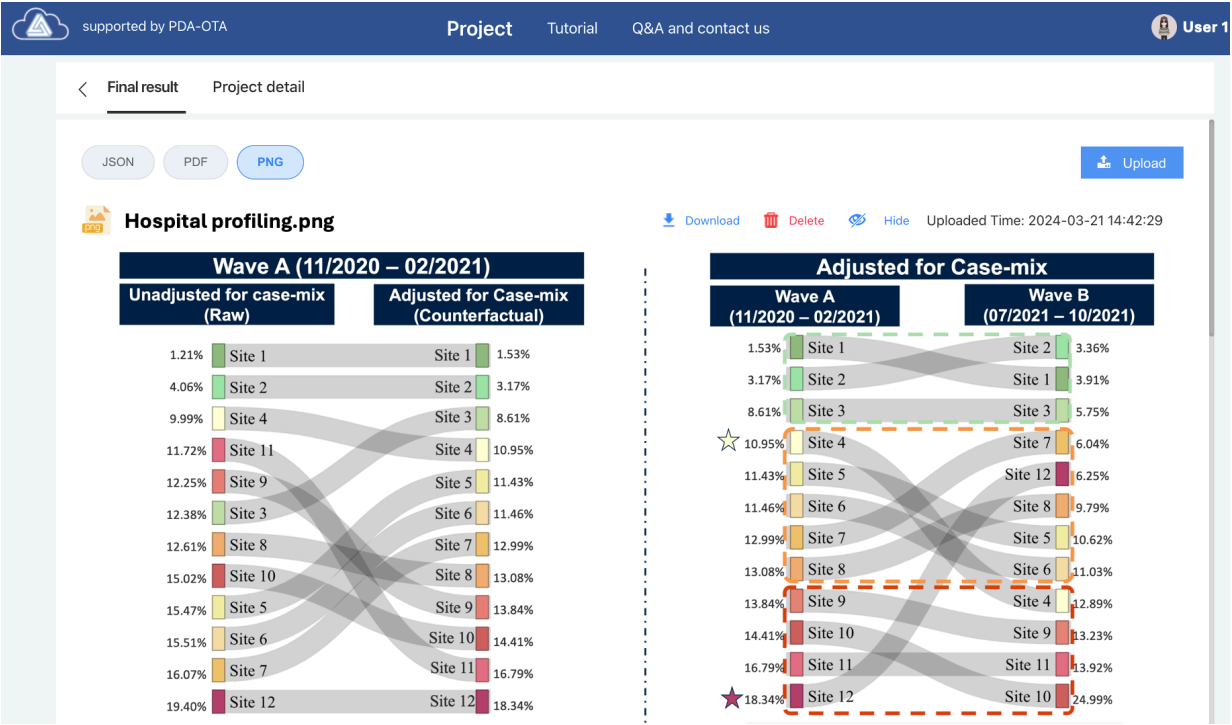

*Figure 4: Example of "Final Result" module within PDA-OTA: result for a PDA study on hospital profiling.*

**DISCUSSION**

In response to the challenges of integrating clinical observational data from multiple databases, this paper provides ten guiding principles for conducting PDA studies aimed at enhancing clinical evidence generation. Adherence to these principles ensures a principled, efficient, collaborative and effective approach to employing distributed learning algorithms within DRNs for clinical evidence generation. Collectively, these principles also support informed decision-making for policymakers and contribute to the advancement of healthcare research.

Guided by these principles, the PDA framework integrates advanced distributed algorithms designed to tackle practical challenges frequently encountered in real-world applications. For instance, it incorporates methodologies capable of addressing various forms of heterogeneity, such as discrepancies in data capture timing and variations in data quality. A key focus of PDA moving forward is the inclusion of more algorithms that exhibit both lossless and one-shot properties — ideal for collaborative federated learning studies. The lossless feature ensures that estimation and prediction accuracy remain identical to analyses using pooled data, with no loss of accuracy due to data-sharing constraints. Meanwhile, one-shot algorithms require only a single round of communication between participating data partners, further enhancing efficiency. There are several one-shot, lossless algorithms with real-world applications that have been developed, and their implementations are publicly available online [23,35,36]. These innovations empower real-time insights in clinical investigations and public health surveillance, improve the efficiency of healthcare services, and accelerate decision-making processes through large-scale collaborative studies.

Looking ahead, we in the PDA framework community are committed to expanding across a wide range of fields to address diverse clinical and regulatory questions as part of its future agenda. To connect the evidence generated through PDA, based on its guiding principles, with its clinical and regulatory applications, the framework acknowledges the critical role of collaborating closely with domain experts and regulatory agencies and engaging in in-depth discussions about the insights derived from PDA studies. The knowledge gained from these multi-site studies should enhance patient-centered outcomes analyses, inform clinical and regulatory decision-making processes, and benefit all stakeholders in the healthcare system

**CONCLUSION**

In this paper, we present ten guiding principles for users to conduct multi-site PDA studies using the algorithms within the PDA framework. Through the strategic synthesis of large-scale clinical observational data from distributed research networks, PDA is dedicated to generating reliable evidence capable of addressing practical research questions or regulatory questions simultaneously. This is achieved with a commitment to transparency, reproducibility, and a systematic approach to data analysis. A number of publications using the PDA framework highlight the successful application of this framework for clinical evidence generation, demonstrating the production of high-quality evidence. The evidence produced using the PDA framework is promising to fill critical gaps in current medical knowledge, offering valuable insights to inform and enhance medical and regulatory decision-making processes.

**FUNDING STATEMENT**

This work was supported in part by National Institutes of Health (U01TR003709, U24MH136069, U24AG098157, RF1AG077820, R01AG073435, R01LM013519, 1R01LM014344, R01DK128237, R21AI167418, R21EY034179).

**AUTHOR CONTRIBUTIONS**

YC and JT conceived and designed the study; YC, JT, and YL developed the visualization framework; JT and YL generated the visualizations; YC and JT drafted the original manuscript; YC, JT, YL, RD, CL, MAS, PBR, AEW, JHH, JHM, HX, YL, RJC, SLZ, GH, and MJS reviewed and edited the manuscript; YC and JT supervised the project, administered the study, and acquired funding. All authors approved the final manuscript.

**CONFLICT OF INTEREST STATEMENT**

None declared.